\documentclass[AER,JEL]{AEA}

\usepackage{natbib}
\usepackage{graphicx}
\usepackage{booktabs}
\usepackage{tikz}
\usepackage{pgfplots}
\pgfplotsset{compat=1.17}
\usetikzlibrary{positioning,fit,backgrounds,calc}

\draftSpacing{1.5}

\begin{document}

\title{The Agentic ETF: How Agentic Trading Becomes an Asset Class}
\shortTitle{The Agentic ETF}
\author{Amandeep Singh\thanks{Scalar Field Research, \texttt{research@scalarfield.io}. This paper develops a conceptual and empirical framework; market-size figures are drawn from public industry data and are attributed at point of use. Forward-looking projections are clearly labeled as such and are not investment advice.}}
\date{\today}
\pubMonth{June}
\pubYear{2026}
\pubVolume{}
\pubIssue{}


\begin{abstract}
Three structural trends are converging in public markets: the exchange-traded fund (ETF) has become the dominant fund wrapper, actively managed ETFs are the fastest-growing segment within it, and algorithmic systems now mediate the overwhelming majority of trading volume. We argue that their intersection defines a nascent asset class---the \emph{Agentic ETF}, a fund whose investment process is delegated to autonomous, large-language-model-driven trading agents that reason over data and execute without a human in the loop. We define agentic trading and distinguish it from rule-based algorithmic trading and robo-advice, decompose the agentic-trading infrastructure into six layers, and identify the incumbent players and structural gaps at each layer. We then present ScalarField.io as a reference implementation of an integrated agentic-trading stack---spanning a unified multi-venue execution registry, isolated strategy agents with continuous broker reconciliation, first-party market data, and metered compute---and map its primitives to each layer of the stack. A transparent, assumption-stated sizing exercise suggests the Agentic ETF could plausibly reach hundreds of billions to low trillions of dollars in assets under management by 2030.
\end{abstract}

\maketitle

The exchange-traded fund is no longer merely a low-cost index vehicle; it has become the default wrapper for accessing public markets. Global ETF assets reached a record US\$19.85 trillion at the end of 2025, a 33.7 percent increase over the US\$14.85 trillion recorded one year earlier, spread across more than 15{,}800 products \citep{etfgi2026total}. Over the same horizon the wrapper itself has displaced competing structures: ETFs grew from roughly 15 percent of total fund assets in 2013 to about 30 percent in 2025 \citep{ishares2025active}. The ETF has won the distribution layer of asset management.

Within that wrapper, the center of gravity is shifting from passive replication to active management. Assets in actively managed ETFs rose 64.5 percent during 2025---from US\$1.17 trillion to US\$1.92 trillion---and reached US\$2.33 trillion by the end of April 2026 \citep{etfgi2026active,etfgi2026activeapr}. Active funds accounted for 88 percent of US ETF launches and 51 percent of global launches through the first half of 2025, and one major issuer projects global active-ETF assets to triple to US\$4.2 trillion by 2030 \citep{blackrock2025active}. The wrapper that won distribution is now absorbing the part of the industry where investment \emph{decisions}, not just exposures, are made.

In parallel, the decision-making process behind trades has been almost fully automated. Industry estimates place the share of global trading volume facilitated by algorithmic systems at roughly 89 percent, of which an estimated 35--45 percent now relies on advanced machine-learning models rather than static rules \citep{tradealgo2026state}. Algorithmic trading is a mature, institution-dominated activity: institutional investors held roughly 61 percent of the algorithmic-trading market in 2025 \citep{mordor2026algo}. A long literature documents that this automation reshaped liquidity and price formation \citep{hendershott2011algo,biais2015hft}.

This paper argues that the intersection of these three trends---the ETF as dominant wrapper, the rise of active ETFs, and the automation of the investment decision---defines an emerging asset class we call the \emph{Agentic ETF}: a fund whose portfolio is managed end-to-end by autonomous trading agents. The recent maturation of large language models (LLMs) capable of tool use and multi-step reasoning \citep{yao2023react,park2023generative}, together with finance-specialized models \citep{wu2023bloomberggpt}, makes it technically feasible to delegate not just execution but \emph{judgment}---data interpretation, security selection, and sizing---to software. What is missing is not model capability but \emph{infrastructure}: the connective tissue that lets an agent observe markets, decide, execute across venues, and have its positions reconciled and accounted for with the reliability a fund requires.

We make four contributions. First, we define agentic trading and separate it cleanly from rule-based algorithmic trading and robo-advice (Section~\ref{sec:agentic}). Second, we characterize the Agentic ETF as an asset class and provide a transparent sizing exercise (Section~\ref{sec:assetclass}). Third, we decompose the agentic-trading infrastructure into six layers and map the incumbents and gaps at each (Section~\ref{sec:stack}). Fourth, we present Scalar Field as a reference implementation that addresses each layer through a single Python interface, and we map its primitives to the stack (Section~\ref{sec:scalarfield}). Section~\ref{sec:risks} discusses risks and regulation, and Section~\ref{sec:conclusion} concludes.

\section{The Convergence Thesis}
\label{sec:convergence}

\begin{figure}[t]
\centering
\begin{tikzpicture}
\begin{axis}[
    ybar,
    width=0.9\textwidth,
    height=6.2cm,
    bar width=22pt,
    ymin=0,
    ymax=4.8,
    ylabel={Active ETF AUM (US\$ trillion)},
    symbolic x coords={2024,2025,Apr-2026,2030 (proj.)},
    xtick=data,
    nodes near coords,
    nodes near coords style={font=\footnotesize},
    enlarge x limits=0.18,
    tick label style={font=\footnotesize},
    label style={font=\footnotesize},
]
\addplot coordinates {(2024,1.17) (2025,1.92) (Apr-2026,2.33) (2030 (proj.),4.2)};
\end{axis}
\end{tikzpicture}
\caption{Actively managed ETF assets under management, actuals and industry projection.}
\begin{figurenotes}[Source]
ETFGI for 2024, 2025, and April 2026 actuals \citep{etfgi2026active,etfgi2026activeapr}; the 2030 figure is a BlackRock projection \citep{blackrock2025active} and is illustrative, not a forecast by the authors.
\end{figurenotes}
\label{fig:active}
\end{figure}
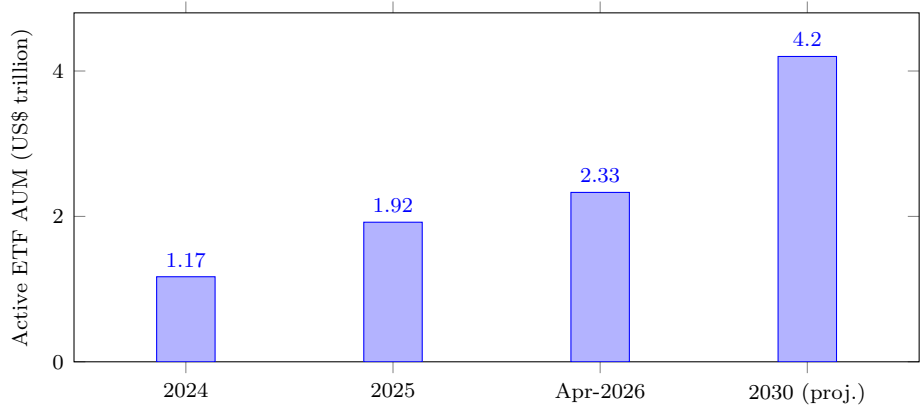

The Agentic ETF is best understood as the meeting point of three independently observable curves. The first is wrapper dominance: capital is consolidating into the ETF structure for its intraday liquidity, tax efficiency, and transparency. The second is the migration of active management into that structure, evidenced by the active-ETF growth in Figure~\ref{fig:active}. The third is the automation of the decision process itself, which has progressed from deterministic execution algorithms toward learning systems.

Crucially, these curves are complementary rather than independent. The ETF wrapper provides distribution and a continuously priced, auditable NAV; active management provides the mandate to make decisions; and automation provides a mechanism to make those decisions at scale and at marginal cost approaching zero. An autonomous agent that can be supervised, throttled, and audited is a natural fit for a wrapper that already demands daily transparency and intraday liquidity. Table~\ref{tab:market} summarizes the magnitudes involved.

\begin{table}[t]
\caption{The three converging markets (most recent available figures).}
\centering
\begin{tabular}{lll}
\toprule
Market & Latest figure & Source \\
\midrule
Global ETF AUM (end-2025) & US\$19.85T (+33.7\% YoY) & \citet{etfgi2026total} \\
Active ETF AUM (end-2025) & US\$1.92T (+64.5\% YoY) & \citet{etfgi2026active} \\
Active ETF AUM (Apr-2026) & US\$2.33T & \citet{etfgi2026activeapr} \\
Active share of US ETF launches (H1-2025) & 88\% & \citet{blackrock2025active} \\
Algorithmic share of global volume & $\approx$89\% & \citet{tradealgo2026state} \\
Institutional share of algo market (2025) & 61\% & \citet{mordor2026algo} \\
ETF share of total fund assets (2025) & $\approx$30\% & \citet{ishares2025active} \\
\bottomrule
\end{tabular}
\begin{tablenotes}
Figures are reported by the cited industry sources and reflect differing methodologies; algorithmic-trading market-size estimates in particular vary widely by scope.
\end{tablenotes}
\label{tab:market}
\end{table}

\section{What Is Agentic Trading?}
\label{sec:agentic}

We define \emph{agentic trading} as a trading process in which an autonomous software agent---typically driven by a large language model with access to tools---perceives market state, reasons over heterogeneous data, forms an investment decision, and executes it across one or more venues, on a recurring schedule, without a human approving each trade. The defining property is the delegation of \emph{judgment}, not merely execution.

This distinguishes agentic trading from two adjacent paradigms. Rule-based algorithmic trading---including execution algorithms such as VWAP and the high-frequency strategies studied by \citet{hendershott2011algo} and \citet{biais2015hft}---automates \emph{how} a predetermined order is worked, but the decision of \emph{what} and \emph{whether} to trade is supplied externally by a human or a fixed model. Robo-advice automates portfolio construction but typically constrains itself to static allocations across passive funds and rebalances mechanically; it does not interpret news, earnings, or cross-asset signals to form discretionary views. Agentic trading sits above both: the agent decides what to trade and why, can revise its thesis as conditions change, and treats execution algorithms and venues as tools it invokes.

Three properties make an agent suitable to manage capital rather than merely to chat. First, \emph{isolation}: each agent must operate against its own capital and position book so that its behavior is measurable and its risk is contained. Second, \emph{scheduled autonomy}: the agent runs on a cadence and acts within that cadence, rather than waiting for a prompt. Third, \emph{accounting truth}: every action must reduce to a verifiable change in positions and cash, reconciled against the venue of record, so that performance can be reported as a NAV time series exactly as a fund would. These properties are precisely what a fund wrapper requires, and they are what the infrastructure of Section~\ref{sec:stack} must provide.

\section{The Agentic ETF as an Asset Class}
\label{sec:assetclass}

An Agentic ETF is the natural productization of agentic trading: an exchange-listed fund whose stated mandate is implemented by one or more autonomous agents rather than by a human portfolio-management team. The wrapper contributes distribution, daily transparency, tax efficiency, and a continuously observable NAV; the agent contributes the investment process. Because an agent's marginal cost of managing an additional mandate is near zero, the structure supports a long tail of narrow, thematic, or personalized mandates that would be uneconomic under human management.

To bound the opportunity, we use a transparent decomposition rather than a point forecast. Let $E$ denote total ETF assets, $a$ the active share of those assets, and $\pi$ the share of active-ETF assets managed agentically. Then Agentic ETF AUM is
\begin{equation}
\text{AUM}_{\text{agentic}} = E \cdot a \cdot \pi .
\label{eq:tam}
\end{equation}
Using the industry projection of $E \approx \text{US\$}27\text{T}$ by 2030 with an active share of $a \approx 16\%$ (i.e.\ active-ETF AUM of US\$4.2T) \citep{ishares2025active,blackrock2025active}, Equation~\eqref{eq:tam} implies an addressable base of US\$4.2T into which agentic management can penetrate. Applying conservative, moderate, and aggressive agentic-penetration assumptions yields the scenarios in Table~\ref{tab:tam}. Even the conservative scenario---agents managing 5 percent of active-ETF assets---implies roughly US\$210 billion, a figure comparable to a large present-day asset-management franchise; the aggressive scenario approaches the size of the entire active-ETF market as of early 2026.

\begin{table}[t]
\caption{Illustrative Agentic ETF sizing under Equation~\eqref{eq:tam}, 2030 base.}
\centering
\begin{tabular}{lll}
\toprule
Scenario & Agentic penetration $\pi$ & Implied AUM \\
\midrule
Conservative & 5\% & US\$0.21T \\
Moderate & 20\% & US\$0.84T \\
Aggressive & 50\% & US\$2.10T \\
\bottomrule
\end{tabular}
\begin{tablenotes}
Computed as US\$4.2T $\times\,\pi$. The US\$4.2T active-ETF base is a third-party projection \citep{blackrock2025active}, and the penetration rates are illustrative assumptions chosen by the authors to bracket plausible outcomes, not forecasts.
\end{tablenotes}
\label{tab:tam}
\end{table}

The point of the exercise is not the specific figures but the structure of the bet: agentic management does not need to win the whole market to constitute a major asset class. It only needs to capture a fraction of the segment---active ETFs---that is already the fastest-growing part of the fastest-growing wrapper. The binding constraint on that capture is not investor demand or model quality; it is whether the supporting infrastructure can make autonomous agents safe, auditable, and venue-agnostic enough to sit inside a fund.

\section{The Agentic Infrastructure Stack}
\label{sec:stack}

\begin{figure}[t]
\centering
\begin{tikzpicture}[
  layer/.style={draw, rounded corners=2pt, minimum width=8.4cm, minimum height=0.95cm, align=center, font=\footnotesize},
  node distance=2.5mm,
]
\node[layer] (l6) {\textbf{Distribution / Wrapper} \\ fund administration, authorized participants, the ETF listing};
\node[layer, below=of l6] (l5) {\textbf{Risk / Reconciliation / Settlement} \\ position truth, freeze/unfreeze, share claims, NAV};
\node[layer, below=of l5] (l4) {\textbf{Execution / Venue Connectivity} \\ Alpaca, Interactive Brokers, Polymarket, Jupiter / DEXs};
\node[layer, below=of l4] (l3) {\textbf{Orchestration / Compute} \\ schedulers, sandboxes, agent frameworks, metering};
\node[layer, below=of l3] (l2) {\textbf{Reasoning / Model} \\ OpenAI, Anthropic, Google, xAI};
\node[layer, below=of l2] (l1) {\textbf{Data} \\ Bloomberg, ICE, exchange feeds, alternative data};
\begin{scope}[on background layer]
\node[draw, dashed, rounded corners=3pt, fit=(l1)(l6),
      inner sep=4mm, label={[font=\footnotesize\itshape, rotate=90, anchor=south]left:integrated by Scalar Field}] (box) {};
\end{scope}
\end{tikzpicture}
\caption{The six-layer agentic-trading infrastructure stack.}
\begin{figurenotes}
Each layer has mature incumbents, but they are not integrated into a single agent-operable surface. The dashed boundary indicates the layers Scalar Field integrates (Section~\ref{sec:scalarfield}).
\end{figurenotes}
\label{fig:stack}
\end{figure}
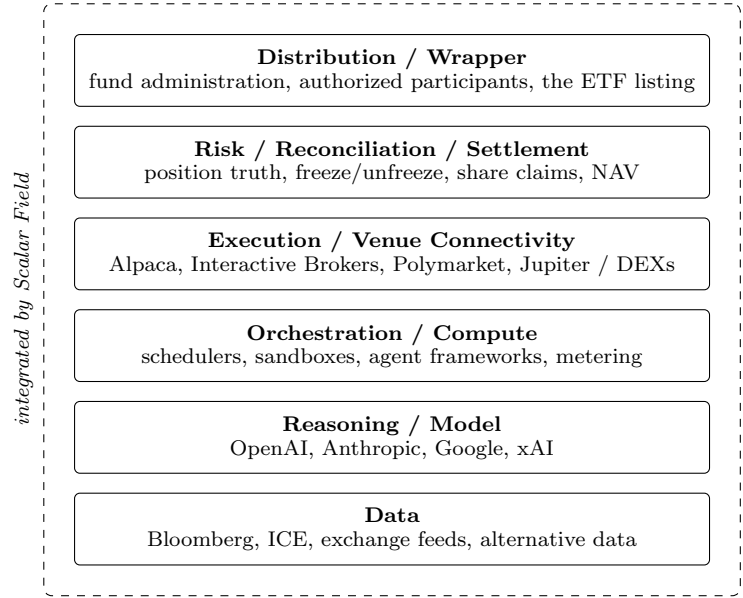

Running an autonomous agent that can be trusted with capital requires a vertically integrated stack. We decompose it into six layers, illustrated in Figure~\ref{fig:stack}. At each layer we identify representative incumbents and the structural gap that an agentic platform must close.

\paragraph{Data layer.} Agents are only as good as what they observe. Incumbents---Bloomberg, ICE, exchange direct feeds, and a fragmented alternative-data market---sell data to humans and institutional pipelines. The gap is agent-readiness: data must be queryable programmatically, normalized across asset classes, and inexpensive enough to consult on every decision cycle. Finance-specialized models such as BloombergGPT illustrate the value of conditioning agents on high-quality financial corpora \citep{wu2023bloomberggpt}.

\paragraph{Reasoning / model layer.} Frontier LLM providers (OpenAI, Anthropic, Google, xAI) supply the reasoning substrate. Tool-use and reason-and-act patterns \citep{yao2023react} make these models capable of multi-step financial workflows. The gap is that a raw model is not a portfolio manager: it has no memory of positions, no execution rights, and no accountability. Reasoning is necessary but far from sufficient.

\paragraph{Orchestration / compute layer.} An agent must run on a schedule, inside a sandbox, with its code, state, and external calls metered and billed. General-purpose agent frameworks exist but are not built for the adversarial, money-moving setting of trading, where a runaway loop or an unmetered model call has direct financial cost. The gap is a trading-grade runtime with deterministic scheduling and cost accounting.

\paragraph{Execution / venue connectivity layer.} Capital must reach markets. Brokers and venues---Alpaca and Interactive Brokers in equities, Polymarket in prediction markets, Jupiter and other DEXs in tokenized assets---each expose idiosyncratic APIs, order types, and settlement models. The gap is a uniform execution surface that lets a single agent trade equities, prediction markets, and tokenized assets without bespoke integration per venue.

\paragraph{Risk / reconciliation / settlement layer.} This is the layer that turns an agent into a fund. The platform must hold the authoritative position book, reconcile it continuously against each venue's actual holdings, detect drift, and halt trading when the book and the broker disagree. It must arbitrate when multiple agents share one brokerage account, tracking which shares each agent has claimed. Generic agent infrastructure ignores this layer entirely; without it, agentic trading cannot be audited and therefore cannot be wrapped in a fund.

\paragraph{Distribution / wrapper layer.} Finally, the strategy must be packaged for investors---fund administration, authorized participants, and the exchange listing that makes it an ETF. This layer is well served by incumbents for traditional funds; the gap is connecting it to an agentic engine whose NAV and holdings are produced by software rather than by a back office.

The thesis of this section is that every layer has capable incumbents, but they are not composed into a single surface an agent can operate. The value is in the integration.

\section{How Scalar Field Solves It}
\label{sec:scalarfield}

Scalar Field is a quantitative research platform that provides the compute, data, and infrastructure to go from research to live trading through a single Python library, \texttt{scalarlib} \citep{scalarfield2026docs}. We treat it as a reference implementation of the integrated stack of Section~\ref{sec:stack}; Table~\ref{tab:mapping} maps each layer to the corresponding primitive.

\paragraph{Unified execution across asset classes.} The \texttt{venue} module exposes one interface---\texttt{account()}, \texttt{positions()}, \texttt{balances()}, \texttt{history()}, and \texttt{trade()}---across a registry that spans US equities (Alpaca), prediction markets (Polymarket), and tokenized assets on Solana (Jupiter DEX), with read-only links to Robinhood, Interactive Brokers, and Fidelity \citep{scalarfield2026docs}. Venues are organized into compatibility groups, so a strategy written for one equities venue can migrate to another without code changes. This collapses the execution/connectivity layer into a single call surface.

\paragraph{Strategy agents with isolated books.} The \texttt{strategy} module is the productized agent. Each strategy receives its own cash allocation, maintains a position book that is isolated from sibling strategies on the same brokerage account, and tracks its own NAV over time. Orders are placed by \emph{target} position---\texttt{execute("AAPL", 10)} declares the desired holding and the platform computes the delta---which makes execution idempotent and therefore safe to retry. This directly implements the isolation, scheduled autonomy, and accounting-truth properties of Section~\ref{sec:agentic}.

\paragraph{Continuous reconciliation as the safety layer.} Scalar Field maintains the authoritative ledger and reconciles the aggregate of all strategy positions against the broker's actual holdings. If a strategy's protected holdings fall below the reconciled floor, it is automatically \emph{frozen} and refuses new orders until realigned; when multiple strategies share an account, the platform tracks which shares each has claimed and rejects seeding that would double-count them \citep{scalarfield2026docs}. This is exactly the risk/reconciliation/settlement layer that converts an autonomous agent into something a fund can rely upon.

\paragraph{First-party data and in-agent reasoning.} The platform ships more than a dozen normalized datasets---equity, index, crypto, forex, and tokenized OHLCV; options quotes and Greeks; insider, congressional, and institutional filings; and prediction-market data---queryable from the same library. Agents reason with \texttt{query\_llm()}, which routes to a menu of models (from Gemini Flash to Claude and GPT) priced by a transparent token-times-multiplier formula. This addresses the data and reasoning layers and meters them at the orchestration layer.

\paragraph{Metered compute.} Execution time, model calls, data services, and alerts are billed through a single credit system---agent execution at a fixed rate per second, LLM calls by tokens and model multiplier \citep{scalarfield2026docs}. Cost accounting is therefore native to the runtime, which is the missing piece at the orchestration/compute layer.

\begin{table}[t]
\caption{Mapping the infrastructure stack to Scalar Field primitives.}
\centering
\begin{tabular}{lll}
\toprule
Layer & Representative incumbents & Scalar Field primitive \\
\midrule
Data & Bloomberg, ICE, exchange feeds & 12+ first-party \texttt{scalarlib} datasets \\
Reasoning / model & OpenAI, Anthropic, Google, xAI & \texttt{query\_llm()} model router \\
Orchestration / compute & agent frameworks, schedulers & scheduled agents + credit metering \\
Execution / connectivity & Alpaca, IBKR, Polymarket, Jupiter & \texttt{venue} registry, \texttt{strategy.execute()} \\
Risk / reconciliation & (largely absent) & isolated books, freeze, share claims \\
Distribution / wrapper & fund admin, APs, exchanges & NAV time series, auditable ledger \\
\bottomrule
\end{tabular}
\begin{tablenotes}[Source]
Scalar Field documentation \citep{scalarfield2026docs}. The wrapper layer is partially addressed: Scalar Field produces the fund-grade NAV and ledger that a listed Agentic ETF would require, while exchange listing and fund administration remain external.
\end{tablenotes}
\label{tab:mapping}
\end{table}

By collapsing the six layers into one programmable surface with reconciliation as a first-class concern, Scalar Field lowers the marginal cost of standing up an auditable autonomous strategy toward the cost of writing the strategy itself---which is the precondition for the long tail of mandates that Section~\ref{sec:assetclass} argues the Agentic ETF will populate.

\section{Risks and Regulation}
\label{sec:risks}

The thesis is not without hazards. \emph{Model risk} is foremost: an LLM can hallucinate a rationale, misread data, or pursue a degenerate objective, and unlike a deterministic algorithm its failure modes are harder to enumerate ex ante. Target-based, idempotent execution and hard reconciliation floors mitigate but do not eliminate this. \emph{Correlated behavior} is a systemic concern: if many agents are conditioned on similar models and signals, they may crowd into the same trades and amplify volatility, echoing the fast-trading externalities formalized by \citet{biais2015hft}. \emph{Execution risk}---slippage, partial fills, and settlement delay on on-chain venues---must be budgeted explicitly in agent code.

Regulation is tightening in parallel. The European Union's AI Act classifies certain financial-sector AI applications as high-risk and phases in obligations on transparency, human oversight, and risk management \citep{euaiact2024}. In the United States, equity market-structure rules continue to govern automated order handling and best execution \citep{secregnms}. An Agentic ETF would additionally sit inside the existing fund-regulatory perimeter, requiring disclosure of the agentic process and controls. The reconciliation and audit layer emphasized above is therefore not only an engineering necessity but a likely regulatory one: an agent that cannot prove what it did, and be halted when it drifts, will struggle to be wrapped in a registered product.

\emph{Limitations of this paper.} Our sizing is a deliberately simple decomposition built on third-party projections, not an econometric forecast; the penetration rates in Table~\ref{tab:tam} are assumptions. We do not present realized returns from a live Agentic ETF, and the claim is structural---about feasibility and addressable market---rather than about alpha. Whether autonomous agents can persistently outperform net of fees is an empirical question we leave to future work, ideally on live, reconciled NAV series of the kind Scalar Field is designed to produce.

\section{Conclusion}
\label{sec:conclusion}

The ETF won distribution; active management is migrating into the wrapper; and the investment decision has been automated. Their intersection is the Agentic ETF---a fund managed end-to-end by autonomous agents. The barrier is not model capability but integrated, fund-grade infrastructure: a single surface where an agent can observe data, reason, execute across venues, and be reconciled and audited. We decomposed that infrastructure into six layers, showed that incumbents serve each layer in isolation but none composes them, and presented Scalar Field as a reference implementation that does. If the supporting stack matures, even modest penetration of the active-ETF market would make the Agentic ETF a multi-hundred-billion-dollar asset class within the decade.

\bibliographystyle{aea}
\bibliography{references}

\end{document}